\documentclass[pra,aps,reprint,superscriptaddress,amsmath]{revtex4-1}

\usepackage{bm}
\usepackage{graphicx}
\usepackage{dsfont}

\newcommand{\bra}[1]{\langle #1 | \,}
\newcommand{\ket}[1]{\, | #1 \rangle}

\newcommand{\bos}[1]{\boldsymbol{#1}}

\newcommand{\mc}[1]{\mathcal{#1}}

\newcommand{\om}{\omega}
\newcommand{\Om}{\Omega}
\newcommand{\ga}{\gamma}
\newcommand{\Ga}{\Gamma}
\newcommand{\de}{\delta}
\newcommand{\De}{\Delta}

\newcommand{\la}{\lambda}
\newcommand{\eps}{\epsilon}
\newcommand{\veps}{\varepsilon}
\newcommand{\hlf}{\frac{1}{2}}
\newcommand{\hc}{\hat{c}}
\newcommand{\ha}{\hat{a}}
\newcommand{\hsig}{\hat{\sigma}}
\newcommand{\teta}{\tilde{\eta}}
\newcommand{\tde}{\tilde{\delta}}

\begin{document}

\title{Microwave to optical conversion with atoms on a superconducting chip}

\author{David Petrosyan}
%\email{david.petrosyan@iesl.forth.gr}
\affiliation{Institute of Electronic Structure and Laser, FORTH,
GR-71110 Heraklion, Crete, Greece}
\affiliation{Department of Physics and Astronomy, Aarhus University,
DK-8000 Aarhus C, Denmark}
\affiliation{Physikalisches Institut, Eberhard Karls Universit\"at T\"ubingen, 
D-72076 T\"ubingen, Germany}

\author{Klaus M\o lmer}
\affiliation{Department of Physics and Astronomy, Aarhus University,
DK-8000 Aarhus C, Denmark}

\author{J\'ozsef Fort\'agh}
\affiliation{Physikalisches Institut, Eberhard Karls Universit\"at T\"ubingen, 
D-72076 T\"ubingen, Germany}

\author{Mark Saffman}
\affiliation{Department of Physics, University of Wisconsin-Madison,
Madison, Wisconsin 53706, USA}

\date{\today}

\begin{abstract}
We describe a scheme to coherently convert a microwave photon 
of a superconducting co-planar waveguide resonator 
to an optical photon emitted into a well-defined temporal and spatial mode. 
The conversion is realized by a cold atomic ensemble trapped close the surface 
of the superconducting atom chip, near the antinode of the microwave cavity. 
The microwave photon couples to a strong Rydberg transition of the atoms 
that are also driven by a pair of laser fields with appropriate frequencies 
and wavevectors for an efficient wave-mixing process. 
With only several thousand atoms in an ensemble of moderate density, 
the microwave photon can be completely converted into an optical photon 
emitted with high probability into the phase matched direction
and, e.g., fed into a fiber waveguide. 
This scheme operates in a free-space configuration, without requiring 
strong coupling of the atoms to a resonant optical cavity.
\end{abstract}

\maketitle

\section{Introduction}

Superconducting quantum circuits, which operate in the microwave frequency 
range, are promising systems for quantum information processing 
\cite{Clarke2008,Devoret2013}, as attested by the immense recent 
interest of academia and industry.
%(e.g. Google, Microsoft, Intel, IBM, 
%D-Wave Systems Inc., Rigetti Computing, Quantum Circuits Inc.). 
On the other hand, photons in the optical and telecommunication
frequency range are the best and fastest carriers of quantum information 
over long distances \cite{Kimble2008,OBrien2009}. 
Hence there is an urgent need for efficient, coherent and reversible conversion 
between microwave and optical signals at the single quantum level
\cite{Kurizki2015}. 
Here we describe such a scheme, which is compatible with both superconducting 
quantum information processing and optical quantum communication technologies.  

Previous work on the microwave to optical conversion includes studies of 
optically active dopants in solids \cite{OBrien2014,Williamson2014}, as well as
electro-optical \cite{Rueda16} and opto-mechanical \cite{Andrews14} systems. 
Cold atomic systems, however, have unique advantages over the other approaches.
Atomic (spin) ensembles can couple to superconducting microwave resonators 
to realize quantum memory in the long-lived hyperfine manifold of levels 
\cite{Verdu09, Hattermann2017}.
Using stimulated Raman techniques \cite{EITrev2005,Hammerer2010},
such spin-wave excitations stored in the hyperfine transition 
can be reversibly converted into optical photons.
Here we propose and analyze an efficient wave-mixing scheme for 
microwave to optical conversion on a integrated superconducting atom chip.
In our setup, the microwave photon is confined in a coplanar waveguide 
resonator, while a cold atomic ensemble is trapped near the antinode 
of the microwave cavity mode at a distance of several tens of microns 
from the surface of the atom chip. We employ a Rydberg transition between 
the atomic states that strongly couple to the microwave cavity field 
\cite{Petrosyan09, Hogan2012, Hermann-Avigliano2014, Teixeira2015}.
The coupling strength of the atoms to the evanescent field of the cavity 
depends on the atomic position, while the proximity of the atoms to the 
chip surface leads to inhomogeneous Rydberg level shifts and thereby 
position-dependent detuning of the atomic resonance. This reduces the 
effective number of atoms participating in four-wave mixing 
in the presence of a pair of laser fields with appropriate frequencies 
and wavevectors. Nevertheless, we show that high-efficiency conversion 
of a microwave photon to an optical photon emitted into a well-defined
spatial and temporal mode is still possible in this setup. The coplanar 
waveguide resonator can also contain superconducting qubits, and hence 
our scheme can serve to interface them with optical photons. 

We note a related work \cite{Han2018} on microwave to optical conversion 
using free-space six-wave mixing involving Rydberg states. The achieved 
photon conversion efficiency was, however, low, as only a small portion 
of the free-space microwave field could interact with the active atomic 
medium. Confining the microwave field in a cavity would be a valuable 
route to enhance the conversion efficiency. 
A microwave to optical conversion scheme using a single (Cs) atom that 
interacts with a superconducting microwave resonator on the Rydberg 
transition and with an optical cavity on an optical transition was 
discussed in \cite{Gard2017}. The advantage of the single atom approach 
is that it requires moderate laser power for atom trapping and leads 
to less light scattering and perturbation of the superconducting resonator. 
It relies, however, on the technically demanding strong coupling 
of the single atom to both microwave and optical cavities. 
Reference \cite{Covey2019} discusses the conversion of a microwave 
photon to an optical telecommunication (E-band) photon employing 
four-wave mixing in a small ensemble of (Yb) atoms in a copper 
microwave resonator and a high-finesse optical cavity. 
In contrast, our present approach uses a large ensemble of atoms 
with collectively enhanced coupling to the microwave cavity and
it leads to a coherent, directional emission of the optical photon 
even without an optical cavity. 
In a previous publication \cite{Petrosyan2018}, we have employed 
a similar scheme to deterministically produce single photons from 
a Rydberg excitation of a single source atom coupled to the atomic 
ensemble via resonant dipole-dipole interaction.

Our setup is primarily intended for optical communication between 
microwave operated quantum sub-registers. As such, we consider the 
case of at most one microwave photon encoding a qubit state at a time. 
The conversion of a microwave photon is accompanied by a Rydberg excitation 
of the atomic ensemble. But since at most only a single atom is excited 
to the Rydberg state, the interatomic interactions and the resulting 
Rydberg excitation blockade \cite{Lukin2001,rydQIrev} do not play a role 
in our scheme, irrespective of whether the atomic ensemble is larger or 
not than any (irrelevant) blockade distance. This allows us to restrict 
the analysis to the linear regime of conversion, greatly simplifying 
the corresponding calculations presented below. 

\section{The system}

%%%%%%%%%%FIGURE%%%%%%%%%%%%
\begin{figure}[t]
%\centerline{\includegraphics[width=8.7cm]{schemeSC.pdf}}
\centerline{\includegraphics[width=8.7cm]{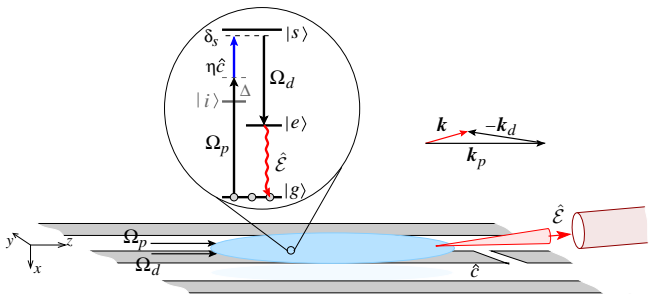}}
\caption{%(color online)
Schematics of the system:  
An ensemble of atoms trapped on a superconducting chip near 
a coplanar waveguide cavity converts the microwave photon 
of the cavity to an optical photon fed into a fiber waveguide, 
as shown in the lower part of the figure.
The inset shows the atomic level scheme.
All the atoms are initially in the ground state $\ket{g}$.
A laser pulse couples $\ket{g}$ to the intermediate Rydberg state $\ket{i}$ 
with the Rabi frequency $\Omega_p$ and detuning $\Delta_p \simeq \Delta$. 
The microwave cavity mode $\hc$ is coupled non-resonantly to the Rydberg 
transition $\ket{i} \to \ket{s}$ of the atoms with a position-dependent
coupling strength $\eta$ and detuning $\Delta_c \simeq -\Delta$. 
With large one-photon detunings $|\Delta_{p,c}| \gg |\Omega_p|,\eta$,
the two-photon transition $\ket{g} \to \ket{s}$ to the Rydberg 
state $\ket{s}$ is detuned by $\delta_s = \Delta_p + \Delta_c$.
A strong laser field $\Omega_d$ drives the transition from $\ket{s}$ 
to the electronically excited state $\ket{e}$ that rapidly decays with rate 
$\Gamma_e > \Omega_d$ to the ground state $\ket{g}$ and emits 
a photon $\mc{E}$ predominantly into the phase-matched direction 
determined by the wave vector $\bos{k} = \bos{k}_p - \bos{k}_d$.}
\label{fig:scheme}
\end{figure}
%%%%%%%%%%%%%%%%%%%%%%%%%%%%%%%%%%%%%%%

Consider the system shown schematically in Fig.~\ref{fig:scheme}. 
An integrated superconducting atom chip incorporates a microwave resonator, 
possibly containing superconducting qubits, 
and wires for magnetic trapping of the atoms. 
An ensemble of $N \gg 1$ cold atoms is trapped near the chip surface, 
close to the antinode of the microwave cavity field. The relevant states 
of the atoms are the ground state $\ket{g}$, a lower electronically excited 
state $\ket{e}$ and a pair of highly-excited Rydberg states $\ket{i}$ and 
$\ket{s}$ (see the inset of Fig.~\ref{fig:scheme}). 
A laser field of frequency $\omega_p$ couples the ground state $\ket{g}$ 
to the Rydberg state $\ket{i}$ with time-dependent Rabi frequency $\Omega_p$ 
and large detuning $\Delta_p \equiv \omega_p - \omega_{ig} \gg |\Omega_p|$.
The atoms interact non-resonantly with the microwave cavity mode $\hc$ 
on the strong dipole-allowed transition between 
the Rydberg states $\ket{i}$ and $\ket{s}$. 
The corresponding coupling strength (vacuum Rabi frequency) 
$\eta = (\wp_{si}/\hbar) \veps_c u(\bos{r})$ is proportional 
to the dipole moment $\wp_{si}$ of the atomic transition, 
the field per photon $\veps_c$ in the cavity, and the cavity mode function 
$u(\bos{r})$ at the atomic position $\bos{r}$. 
The Rydberg transition is detuned from the cavity mode resonance 
by $\Delta_c \equiv \omega_c - \omega_{si}$, $|\Delta_c| \gg  \eta$. 
A strong driving field of frequency $\omega_d$ acts on the transition 
from the Rydberg state $\ket{s}$ to the lower excited state $\ket{e}$
with Rabi frequency $\Omega_d$ and detuning $\Delta_d = \omega_d - \omega_{se}$. 
The transition from the excited state $\ket{e}$ to the ground state $\ket{g}$ 
is coupled with strengths $g_{\bos{k},\sigma}$ 
%$g_{\bos{k},\sigma} = \frac{\bos{\wp}_{eg} \cdot \bos{e}_{\bos{k},\sigma} }{\hbar} 
%\sqrt{\frac{\hbar \omega_k}{2 \eps_o V}}$ 
to the free-space quantized radiation field modes $\ha_{\bos{k},\sigma}$ 
characterized by the wave vectors $\bos{k}$, polarization $\sigma$ 
and frequencies $\omega_k =c k$. 

In the frame rotating with the frequencies of all the fields, 
$\omega_p$, $\omega_c$, $\omega_d$, and $\omega_k$, 
dropping for simplicity the polarization index, 
the Hamiltonian for the system reads
\begin{eqnarray}
H/\hbar &=& - \sum_{j=1}^N 
\Big[ \Delta_p^{(j)} \hsig_{ii}^{(j)} + \delta_s^{(j)} \hsig_{ss}^{(j)} 
+ \delta_e \hsig_{ee}^{(j)} 
\nonumber \\ & & 
+ \Big( \Omega_p e^{i \bos{k}_p \cdot \bos{r}_j} \hsig_{ig}^{(j)} 
- \eta(\bos{r}_j) \hc \, \hsig_{si}^{(j)}  
+ \Omega_d e^{i \bos{k}_d \cdot \bos{r}_j} \hsig_{se}^{(j)}  
\nonumber \\ & & \;\;
+ \sum_{\bos{k}}  g_{\bos{k}}  \ha_{\bos{k}} 
e^{i \bos{k} \cdot \bos{r}_j} e^{-i (\omega_k - \omega_{eg}) t} \hsig_{eg}^{(j)} 
+ \mathrm{H.c.} \Big) \Big] , \label{eq:Hamfull}
\end{eqnarray}
where index $j$ enumerates the atoms at positions $\bos{r}_j$, 
$\hsig_{\mu \nu}^{(j)} \equiv \ket{\mu}_j \bra{\nu}$ are the atomic 
projection ($\mu = \nu$) or transition ($\mu \neq \nu$) operators, 
$\bos{k}_{p}$ and $\bos{k}_{d}$ are the wave vectors of the corresponding 
laser fields,
$\delta_s^{(j)} \equiv \Delta_p^{(j)} + \Delta_c^{(j)} 
= \omega_p + \omega_{c} - \omega_{sg}^{(j)}$ 
is the two-photon detuning of level $\ket{s}$, and 
$\delta_e \equiv \delta_s^{(j)} - \Delta_d^{(j)} 
= \omega_p + \omega_{c} - \omega_d - \omega_{eg}$ 
is the three-photon detuning of $\ket{e}$. 
The energies of the Rydberg levels $\ket{i},\ket{s}$, 
and thereby the corresponding detunings $\Delta_{p,c}^{(j)}$ and $\delta_s^{(j)}$,
depend on the atomic distance $(x_0-x_j)$ from the chip surface at $x_0$, 
which may contain atomic adsorbates producing an inhomogeneous 
electric field \cite{Tauschinsky2010,Hattermann2012}.  
We neglect the level shift of the lower state $\ket{e}$,
since it is typically less sensitive to the electric fields 
and has a large width $\Gamma_e$ (see below).

We assume that initially all the atoms are prepared in the ground state,
$\ket{G} \equiv \ket{g_1,g_2,\ldots,g_N}$, 
the microwave cavity contains a single photon, $\ket{1_c}$,
and all the free-space optical modes are empty, $\ket{0}$.
We can expand the state vector of the combined system as 
$\ket{\Psi} = b_0 \ket{G} \otimes \ket{1_c} \otimes \ket{0}
+ \sum_{j=1}^N d_j e^{i \bos{k}_p \cdot \bos{r}_j} \ket{i_j} 
\otimes \ket{1_c} \otimes \ket{0} 
+ \sum_{j=1}^N c_j e^{i \bos{k}_p \cdot \bos{r}_j} \ket{s_j} 
\otimes \ket{0_c} \otimes \ket{0}
+ \sum_{j=1}^N b_j e^{i (\bos{k}_p - \bos{k}_d)  \cdot \bos{r}_j} \ket{e_j} 
\otimes \ket{0_c} \otimes \ket{0}
+ \ket{G} \otimes \ket{0_c} \otimes 
\sum_{\bos{k}} a_{\bos{k}} \ket{1_{\bos{k}}}$,
%
%\begin{eqnarray*}
%\ket{\Psi} &=& b_0 \ket{G} \otimes \ket{1_c} \otimes \ket{0}
%\\ & & 
%+ \sum_{j=1}^N d_j e^{i \bos{k}_p \cdot \bos{r}_j} \ket{i_j} 
%\otimes \ket{1_c} \otimes \ket{0} 
%\\ & & 
%+ \sum_{j=1}^N c_j e^{i \bos{k}_p \cdot \bos{r}_j} \ket{s_j} 
%\otimes \ket{0_c} \otimes \ket{0}
%\\ & & 
%+ \sum_{j=1}^N b_j e^{i (\bos{k}_p - \bos{k}_d)  \cdot \bos{r}_j} \ket{e_j} 
%\otimes \ket{0_c} \otimes \ket{0}
%\\ & & 
%+ \ket{G} \otimes \ket{0_c} \otimes 
%\sum_{\bos{k}} a_{\bos{k}} \ket{1_{\bos{k}}},
%\end{eqnarray*}
%
where $\ket{\mu_j} \equiv \ket{g_1,g_2, \ldots, \mu_j, \ldots, g_N}$
denote the single excitation states, $\mu = i,s,e$, 
and $\ket{1_{\bos{k}}} \equiv \hat{a}^{\dagger}_{\bos{k}}  \ket{0}$ denotes 
the state of the radiation field with one photon in mode $\bos{k}$.
The evolution of the state vector is governed by the Schr\"odinger 
equation $\partial_t \ket{\Psi} = -\frac{i}{\hbar} H \ket{\Psi}$ 
with the Hamiltonian (\ref{eq:Hamfull}), which leads to the system
of coupled equations for the slowly-varying in space atomic amplitudes,
\begin{subequations}
\label{eqs:b0djcjbj}
\begin{eqnarray}
\partial_t b_0 &=& i \sum_{j=1}^N \Omega_p^* d_j  , \\
\partial_t d_j &=& i \Delta_p^{(j)} d_j + i \Omega_p b_0 - i \eta^* (\bos{r}_j) c_j , \\
\partial_t c_j &=& i \delta_s^{(j)} c_j  -i \eta(\bos{r}_j) d_j   + i \Omega_d b_j , \\
\partial_t b_j &=& i \delta_e b_j + i \Omega_d^* c_j 
\nonumber \\  & &  
+ i \sum_{\bos{k}} g_{\bos{k}} e^{i (\bos{k} - \bos{k}_p + \bos{k}_d)  \cdot \bos{r}_j} 
a_{\bos{k}} e^{- i (\om_k - \om_{eg}) t},
\end{eqnarray}
\end{subequations}
while the equation for the optical photon amplitudes written in the integral
form is
\begin{equation}
a_{\bos{k}} (t) = i g_{\bos{k}}^* 
\sum_{j} e^{ i (\bos{k}_p - \bos{k}_d  - \bos{k}) \cdot \bos{r}_j}
\int_0^t dt' b_j(t') e^{i (\omega_k - \omega_{eg}) t'} . \label{eq:akint}
\end{equation}
The initial conditions for Eqs. (\ref{eqs:b0djcjbj}), (\ref{eq:akint}) 
are $b_0 (0) = 1$, $b_j (0), c_j (0), d_j (0) = 0 \, \forall \, j$, 
and $a_{\bos{k}} (0) = 0 \, \forall \, \bos{k}$. 

We substitute Eq.~(\ref{eq:akint}) into the equation for atomic 
amplitudes $b_j$, assuming they vary slowly in time, and obtain 
the usual spontaneous decay of the atomic state $\ket{e}$ with rate $\Gamma_e$
and the Lamb shift that can be incorporated into $\omega_{eg}$ 
\cite{ScullyZubary1997}.
We neglect the field--mediated interactions (multiple scattering) between 
the atoms \cite{Lehmberg1970,Thirunamachandran,Miroshnychenko2013},
assuming random atomic positions and sufficiently large mean 
interatomic distance $\bar{r}_{ij} \gtrsim \lambda/2\pi$. 
To avoid the atomic excitation in the absence of a microwave photon in the 
cavity, we assume that the intermediate Rydberg level $\ket{i}$ is strongly 
detuned, 
$\Delta_{p}^{(j)} \simeq - \Delta_{c}^{(j)} \gg |\Om_p|,\eta, |\delta_s^{(j)}|$ 
for all atoms in the ensemble. In addition, we assume that the variation 
of $\Delta_{p}^{(j)}$ ($\Delta_{c}^{(j)}$) across the atomic cloud is small 
compared to its mean value $\Delta$ ($-\Delta$), which presumes small enough 
Rydberg levels shifts in the inhomogeneous electric field. 
We can then adiabatically eliminate the intermediate Rydberg level $\ket{i}$,
obtaining finally
\begin{subequations}
\label{eqs:b0cjbj}
\begin{eqnarray}
\partial_t b_0 &=& i \sum_{j=1}^N \teta_j c_j  , \\
\partial_t c_j &=& (i \tde_s^{(j)} - \Gamma_s/2) c_j  + i \teta_j b_0   + i \Omega_d b_j , \\
\partial_t b_j &=& (i \tde_e - \Gamma_e/2)  b_j + i \Omega_d^* c_j , 
\end{eqnarray}
\end{subequations}
where $\teta_j \equiv \frac{\eta(\bos{r}_j) \Omega_p} {\Delta} \big[ 1 + \frac{\delta_s^{(j)}}{2\De} \big]$
is the second-order coupling between $\ket{g_j} \otimes \ket{1_c}$ and $\ket{s_j} \otimes \ket{0_c}$,
while the second-order level shifts of $\ket{g_j}$ and $\ket{s_j}$ are incorporated into the
detunings $\tde_s^{(j)} \equiv \delta_s^{(j)} + \frac{|\Omega_p|^2 - |\eta(\bos{r}_j)|^2}{\Delta}$
and $\tde_e = \delta_e + \frac{|\Omega_p|^2}{\Delta}$.
We have also included the typically slow decay $\Gamma_s$ of state $\ket{s}$ 
corresponding to the loss of Rydberg atoms 
\cite{Hermann-Avigliano2014,Beterov2009}.

Before presenting the results of numerical simulations, 
we can derive an approximate analytic solution of the above equations 
and discuss its implications. 
We take a time-dependent pump field $\Omega_p(t)$ (and thereby $\teta_j(t)$) 
and a constant driving field $\Om_d < \Gamma_e/2$, which results in 
an effective broadening of the Rydberg state $\ket{s}$ by 
$\gamma = \frac{|\Om_d|^2}{\Ga_e/2}$.
Assuming $\ga \gg \Ga_s/2, \tde_e$, we then obtain 
\begin{subequations}
\label{eqs:b0bj}
\begin{eqnarray}
b_j(t) &=& - \frac{\ga}{\Om_d} \frac{\teta_j(t)}{\ga - i \tde_s^{(j)}} \, b_0(t)  , \\
b_0(t) &=& b_0(0) \exp \left[ -\int_0^t dt' \sum_{j=1}^N \frac{|\teta_j(t')|^2} {\ga - i \tde_s^{(j)}} \right]  . 
\end{eqnarray}
\end{subequations}
Substituting these into Eq.~(\ref{eq:akint}) and separating the temporal and spatial dependence,
we obtain
\begin{equation}
a_{\bos{k}} (t) = -i \frac{\ga}{\Om_d} A_k(t) \times B_{\bos{k}} , \label{eq:akintan}
\end{equation}
where
\begin{subequations}
\begin{eqnarray}
A_k(t) &=& \int_0^t \!\! dt' \Om_p(t') \, e^{i (\omega_k - \omega_{eg}) t'} 
e^{- \beta \int_0^{t'}dt^{\prime \prime} |\Om_p(t^{\prime \prime})|^2} , \quad \label{eq:Akt} \\
B_{\bos{k}} &=& \frac{g_{\bos{k}}^*}{\De} 
\sum_{j=1}^N \frac{\eta(\bos{r}_j)}{\ga - i \tde_s^{(j)}} 
\, e^{ i (\bos{k}_p - \bos{k}_d  - \bos{k}) \cdot \bos{r}_j} , \label{eq:Bk}
\end{eqnarray}
\end{subequations}
with 
$\beta = \frac{1}{\De^2}\sum_{j=1}^{N} \frac{|\eta(\bos{r}_j)|^2} {\ga - i \tde_s^{(j)}}$.

Equation (\ref{eq:Akt}) shows that for a sufficiently smooth envelope 
of the pump field $\Omega_p(t)$, the optical photon is emitted within 
a narrow bandwidth $\beta |\Om_p|^2$ around frequency $\omega_k = \omega_{eg}$, 
which is a manifestation of the energy conservation. 
The temporal profile of the photon field at this frequency is 
$\eps (t) = \partial_t A_{k_0}(t) 
= \Om_p(t) e^{- \beta \int_0^{t} dt^{\prime} |\Om_p(t^{\prime})|^2}$,
where $k_0 = \omega_{eg}/c$.
The envelope of the emitted radiation can be tailored  to the desired 
profile $\eps (t)$  by shaping the pump pulse according to $\Omega_p(t) 
= \eps (t) \big[ 1 - 2 \beta \int_0^t d t' |\eps (t')|^2\big]^{-1/2}$
\cite{Kiilerich2019}, which can facilitate the photon transmission 
and its coherent re-absorption in a reverse process at a distant location
\cite{McKeever2004,Hijlkema2007,Reiserer2015}. Neglecting the photon
dispersion during the propagation from the sending to the receiving node,
and assuming the same or similar physical setup at the receiving node
containing an atomic ensemble driven by a constant field $\Omega_d$,
the complete conversion of the incoming optical photon to the cavity 
microwave photon is achieved by using the receiving laser pulse of the shape 
$\Omega_p(t) = - \eps (t) \big[2 \beta \int_0^t d t' |\eps (t')|^2\big]^{-1/2}$
\cite{Kiilerich2019}. 

The spatial profile of the emitted radiation in Eq.~(\ref{eq:Bk}) is 
determined by the geometry of the atomic cloud, the excitation amplitudes 
of the atoms at different positions, and the phase matching.
We assume an atomic cloud with normal density distribution 
$\rho(\bos{r}) = \rho_0 e^{-x^2/2\sigma_x^2 - y^2/2\sigma_y^2 - z^2/2\sigma_z^2}$
in an elongated harmonic trap, $\sigma_z > \sigma_{x,y}$. 
To maximize the resonant emission at frequency 
$\om_{k_0} = c |\bos{k}_p - \bos{k}_d| = \om_{eg}$
into the phase matched direction 
$\bos{k} = \bos{k}_p - \bos{k}_d$, we assume the (nearly) collinear geometry
$\bos{k}_p, \bos{k}_d \parallel \bos{e}_z$. 
In an ideal case of all the atoms having the same excitation amplitude 
$b_j \propto 1/\sqrt{N} \, \forall \, j$, and hence 
$B_{\bos{k}} \propto \int d^3 r \rho(\bos{r}) e^{i (\bos{k}_p-\bos{k}_d-\bos{k}) \cdot \bos{r}}$, 
the photon would be emitted predominantly into an (elliptic) Gaussian mode 
$\mc{E}(\bos{r}) \propto \sum_{|\bos{k}| =k_0} B_{\bos{k}} e^{i \bos{k} \cdot \bos{r}}$ 
with the waists $w_{0x,0y} = 2 \sigma_{x,y}$, namely
\begin{equation}
\mc{E}(x,y,z) = \left(\frac{2}{\pi w_x w_y} \right)^{1/2} 
e^{ i k_0 (z + x^2/2 q_x^* + y^2/2q_y^*)} ,
\end{equation} 
where $w_{x,y} = w_{0x,0y} 
\Big[1 + \big(\frac{z}{\zeta_{x,y}} \big)^2 \Big]^{1/2}$ and
$q_{x,y} = z - i \zeta_{x,y}$ with $\zeta_{x,y} = \frac{\pi w_{0x,0y}^2}{\la_0}$. 
The corresponding angular spread (divergence) of the beam 
is $\Delta \theta_{x,y} = \frac{\lambda_0}{\pi w_{0x,0y}} 
= \frac{1}{k_0 \sigma_{x,y}}$, which spans the solid angle 
$\Delta \Omega = \pi \Delta \theta_{x} \Delta \theta_{y}$. 
The probability of the phase-matched, cooperative photon emission 
into this solid angle is $P_{\Delta \Omega} \propto N \Delta \Omega$, 
while the probability of spontaneous, uncorrelated photon emission 
into a random direction is $P_{4\pi} \propto 4\pi$. 
With $P_{\Delta \Omega} + P_{4\pi} =1$, we obtain 
$P_{\Delta \Omega} = \frac{N \Delta \Omega}{N \Delta \Omega + 4\pi}$ 
which approaches unity for $N \Delta \Omega \gg 4\pi$ or
$N \gg 4 k_0^2 \sigma_{x} \sigma_{y}$ \cite{Saffman2005}. 
Hence, for the product $N \Delta \Omega$, and thereby $P_{\Delta \Omega}$,
to be large, we should take an elongated atomic cloud with 
large $\sigma_{z}$ (to have many atoms $N$ at a given atom density) 
and small $\sigma_{x}, \sigma_{y}$ (to have large solid angle $\Delta \Omega$). 

In our case, however, not all the atoms participate equally 
in the photon emission, since the atomic amplitudes 
$b_j \propto \frac{\eta(\bos{r}_j)}{\ga - i \tde_s^{(j)}}$ depend strongly
on the distance $(x_0-x_j)$ from the chip surface via both the atom-cavity
coupling strength $\eta(\bos{r}_j) \simeq \eta_0 e^{-(x_0-x_j)/D}$, and, 
more sensitively, the Rydberg state detuning $\tde_s^{(j)} \simeq \alpha x_j$ 
(see below). This detuning results in a phase gradient for the atomic 
amplitudes in the $x$ direction, which will lead to a small inclination 
of $\bos{k}$ with respect to $\bos{k}_p - \bos{k}_d$ in the $x-z$ plane. 
More importantly, for strongly varying detuning, $\alpha > 2 \ga/\sigma_x$, 
only the atoms within a finite-width layer $\Delta x < \sigma_x$ are 
significantly excited to contribute to the photon emission. 
This reduces the cooperativity via $N\to \xi N$ with the effective 
participation fraction $\xi \simeq \frac{\Delta x}{\sigma_x} < 1$, but also 
leads to larger divergence $\Delta \theta_{x} \simeq \frac{1}{k_0 \Delta x}$
in the $x-z$ plane.
%as compared to the divergence $\Delta \theta_{y} = \frac{1}{k_0 \sigma_y}$ in the $y-z$ plane. 

\section{Results}

\begin{figure}[t]
%\centerline{\includegraphics[width=8.7cm]{combfig.pdf}}
\centerline{\includegraphics[width=8.7cm]{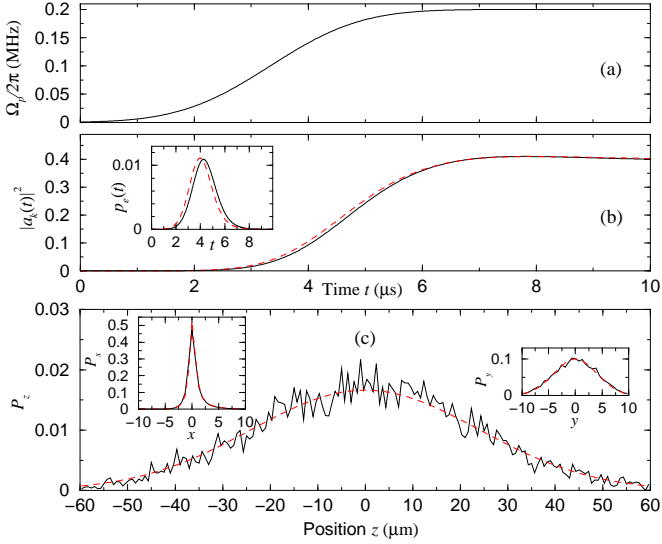}}
\caption{%(color online)
(a) Rabi frequency of the pump field. 
(b) Probability $|a_{\bos{k}} (t)|^2$ (unnormalized) of photon emission 
into the resonant $\om_k = \om_{eg}$, 
phase-matched $\bos{k} = \bos{k}_p - \bos{k}_d$ mode, 
obtained from the exact numerical (black solid line) and analytical 
(red dashed line) solutions. The inset shows the total population
$p_e(t)$ of the atoms in state $\ket{e}$.   
(c) Spatial distribution of time-integrated photon emission probability 
(in any direction) $P$, as obtained from a single realization 
of the atomic ensemble. Main panel shows $P_z$ along the $z$ axis 
(integrated over the $x$ and $y$ direction), while the insets show
$P_x$ and $P_y$ along $x$ and $y$ (black solid lines). 
For comparison, we show the Gaussians 
$P_z = \frac{1}{\sqrt{2 \pi} \sigma_z} e^{-z^2/2 \sigma_z^2}$ and 
$P_y = \frac{1}{\sqrt{2 \pi} \sigma_y} e^{-y^2/2 \sigma_y^2}$ for 
the $z$ and $y$ directions, and 
$P_x = \frac{1}{\mathcal{N}} \frac{|\eta(x)|^2}{\ga^2 + \de_s^2(x)}
e^{-x^2/2 \sigma_x^2}$ along $x$, with $\mathcal{N}$ the normalization
(red dashed lines). }
\label{fig:comb}
\end{figure}

We have verified these arguments via exact numerical simulations of
the dynamics of the system. 
We place $N$ ground state $\ket{g_j}$ atoms in an elongated volume
at random positions $\bos{r}_j$ normally distributed around the origin, 
$x,y,z=0$, with standard deviations $\sigma_{z} \gg \sigma_{x,y}$. 
With the peak density $\rho_0 = 2.35\:\mu$m$^{-3}$ and $\sigma_{x,y} = 4\:\mu$m,
$\sigma_{z} = 24\:\mu$m, we have $N=15000$ atoms in the trap interacting 
with the co-planar waveguide resonator at position $x_0 \simeq 40\:\mu$m
(see Fig.~\ref{fig:scheme}).
Taking the strip-line length $L=10.5\:$mm and the grounded electrodes 
at distance $D=10\:\mu$m, the effective cavity volume is 
$V_c \simeq 2 \pi D^2 L$ \cite{Verdu09} yielding the field per photon 
$\veps_c = \sqrt{\hbar \om_c/\eps_0 V_c } \simeq 0.37\:$V/m
for the full-wavelength cavity mode of frequency 
$\om_c/2\pi = c/L\sqrt{\eps_r} \simeq 12\:$GHz ($\eps_r \simeq 5.6$). 
The atomic cloud is near the antinode of the standing-wave cavity field 
which falls off evanescently with the distance from the chip surface 
as $u(\bos{r}) \simeq e^{-(x_0 - x)/D}$.
The cavity field varies very little along the longitudinal $z$ direction
of the atomic cloud since the cloud dimension $\sigma_{z}$ is much smaller
than the wavelength of the microwave radiation $\la_c = L$.  
We choose the Rydberg states $\ket{i} = \ket{68P_{3/2},m_J=1/2}$ 
and $\ket{s} = \ket{69S_{1/2},m_J=1/2}$ 
of Rb with the quantum defects $\de_P = 2.651$ and $\de_S = 3.131$ 
\cite{RydAtoms}, leading to the transition frequency 
$\om_{si}/2\pi \simeq 12.1\:$GHz and dipole moment $\wp_{si} \simeq 2185a_0 e$. 
This results in the vacuum Rabi frequency $\eta(0)/2\pi \simeq 190\:$kHz 
at the cloud center $\bos{r} = 0$. We take a sufficiently large 
intermediate state detuning $\Delta/2\pi \simeq 10\:$MHz, 
and time-dependent pump field $\Omega_p(t) = \Om_0 \hlf \big[1 
+ \mathrm{erf} \big( \frac{t-t_0}{\sqrt{2} \sigma_t} \big) \big]$ 
of duration $t_{\mathrm{end}} \simeq 10\:\mu$s with $t_0 = t_{\mathrm{end}}/3$, 
$\sigma_t = t_{\mathrm{end}}/8$ and the peak value $\Om_0/2 \pi \simeq 200\:$kHz
(see Fig.~\ref{fig:comb}(a)). The wavelength of the pump field is
$\la_p \simeq 297\:$nm corresponding to a single-photon transition
from $\ket{g} = \ket{5S_{1/2},F=2,m_F =2}$ to $\ket{i}$; 
alternatively, a three-photon transition between states $\ket{g}$ and $\ket{i}$
via the intermediate $5P_{1/2}$ and $6S_{1/2}$ states is possible.
The single photon transition $\ket{g} \to \ket{i}$ has a small dipole 
moment $\wp_{gi} = 2.23 \times 10^{-4}a_0 e$, and the required peak intensity 
of the UV field to attain the Rabi frequency $\Om_0$ is 
$I_0 = \frac{\eps_0 c}{2} \frac{\hbar^2 \Om_0^2}{\wp_{gi}^2} = 650\;$W/cm$^{2}$,
which can be delivered by a laser pulse of $1.5\:$nJ energy focused to a 
spot size of $w_p \simeq 5\:\mu$m.  
The two-photon detuning $\tde_s$ of the Rydberg level $\ket{s}$ 
is taken to be zero at the cloud center $\bos{r} = 0$ 
and it varies with the atomic position along the $x$ axis as 
$\tde_s(x) = \alpha x$ with $\alpha = 2\pi \times 0.5\:$MHz $\mu$m$^{-1}$ 
due to the residual or uncompensated surface charges on the atom chip
\cite{Sedlacek2016,Booth2018}.  
The strong laser field with wavelength $\la_d = 480\:$nm is driving
the transition from the Rydberg state $\ket{s}$ to 
$\ket{e} = \ket{5P_{3/2},F=2,m_F =2}$ with a constant Rabi frequency 
$\Om_d/2\pi = 1\:$MHz. The calculated dipole moment for this
transition is $\wp_{gi} = 3.88 \times 10^{-3}a_0 e$ and the required 
intensity of the driving field is $I_d = 55\;$W/cm$^{2}$. 
The decay rates of states $\ket{s}$ ($\ket{i}$) and $\ket{e}$ are 
$\Ga_{s(i)}/2\pi = 1.6\:$kHz and $\Ga_e/2\pi = 6.1\:$MHz.

We can estimate the absorption of the laser fields by the superconducting
electrodes of the microwave resonator, which would break Cooper pairs and 
reduce the cavity quality factor resulting in microwave absorption.
The intensity of the focused UV pulse at the chip surface is reduced by 
a factor of $e^{-16}$ from its peak value at the cloud center, which means 
that about $3000$ photons will hit the chip surface above the atomic cloud. 
In addition, the atoms in the cloud will scatter the UV photons 
in all $4\pi$ direction with the rate $N \Ga_i (\Om_p/\Delta)^2$, 
but this leads to only $3 \times 10^{-5}$ scattered photons per pulse. 
Assuming the surface reflectivity of $0.999$ (the field propagation 
direction is parallel to the surface), we have only a few absorbed 
photons per pulse, which is negligible compared to the cooling rate 
of the cryogenic environment. 
Similar estimates for the driving field show that only about $700$ 
photons will hit the surface of the atom chip, and less than one will 
be absorbed during the conversion cycle, while the scattering from 
the cloud is negligible since at most only one atom can be excited 
to state $\ket{s}$ at a time.   

%%%%%%%%%%FIGURE%%%%%%%%%%%%
%%%%%%%%%%%%%%%%%%%%%%%%%%%%%%%%%%%%%%%

In Fig.~\ref{fig:comb} we show the results of our numerical simulations 
of the dynamics of the system and compare them with the analytical solutions.
In the inset of Fig.~\ref{fig:comb}(b) we show the time dependence 
of the total population $p_e(t) = \sum_{j=1}^N |b_j(t)|^2$ of the atoms 
in the excited state $\ket{e}$.
As atoms decay from state $\ket{e}$ to the ground state $\ket{g}$, 
they emit a photon with rate $\Ga_e |b_j(t)|^2$. 
The spatial distribution of time-integrated photon emission probability 
(in any direction) $P(\bos{r}_j) = \Ga_e \int_0^{t_{\mathrm{end}}} |b_j(t)|^2 dt$
is shown in Fig.~\ref{fig:comb}(c). This probability follows the Gaussian 
density profile of the atoms along the $y$ and $z$ directions,
but in the $x$ direction it is modified by an approximate Lorentzian factor 
$\frac{|\eta(x)|^2}{\ga^2 + \de_s^2(x)}$ (if we neglect the $x$ dependence of 
$\eta(x)$) due to the position-dependent detuning $\de_s(x)$. 
Only part of the radiation is coherently emitted into the phase-matched
direction $\bos{k} = \bos{k}_p - \bos{k}_d$, with probability $|a_{\bos{k}} (t)|^2$
of photon emission into the resonant $\om_k = \om_{eg}$ mode shown in
Fig.~\ref{fig:comb}(b). Note that for non-resonant modes $\om_k \neq \om_{eg}$
with the rapidly oscillating phase factor $e^{i (\omega_k - \omega_{eg}) t'}$ 
in Eq.~(\ref{eq:akint}) or (\ref{eq:Akt}),
the photon amplitude $a_{\bos{k}} (t) \propto \sum_j b_j(t)$ tends 
to zero at large times $t_{\mathrm{end}}$ (as do $b_j(t)$'s), 
even for the phase-matched direction $\bos{k} \simeq \bos{k}_p - \bos{k}_d$.
 
%%%%%%%%%%FIGURE%%%%%%%%%%%%
\begin{figure}[t]
%\centerline{\includegraphics[width=8.7cm]{pkpcomb.pdf}}
\centerline{\includegraphics[width=8.7cm]{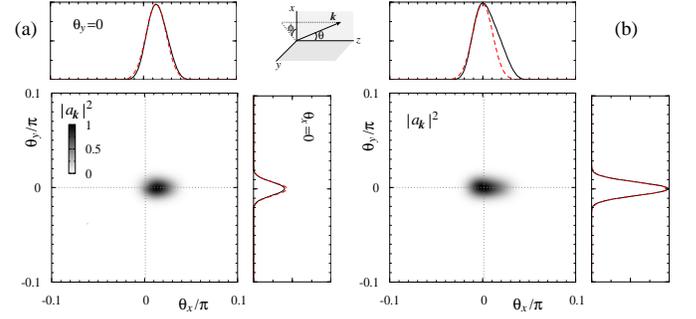}}
\caption{%(color online)
Angular probability distribution $|a_{\bos{k}}|^2$ (unnormalized)
of the photon emitted along the $z$ direction, as a function of
$\theta_x = \theta \cos(\phi)$ and $\theta_y = \theta \sin(\phi)$
with $\theta$ the polar and $\phi$ the azimuthal angles, as shown 
in middle top inset. 
Panel (a) corresponds to the case of the collinear geometry 
$\bos{k}_p, \bos{k}_d \parallel \hat{z}$, with the radiation emitted at a small 
angle $\theta_{x0} = \bos{k} \protect\angle \hat{z} \simeq 0.014 \pi$. 
Panel (b) shows the case with a small inclination 
$\bos{k}_d \protect\angle \hat{z} = 0.009 \pi$ and 
$\bos{k}_p \parallel \hat{z}$, leading to $\bos{k} \parallel \hat{z}$
($\theta_{x0} =0$).
The red dashed lines in the insets of each density plot show 
the Gaussian $B(\theta_x,\theta_{x0};\theta_y,\theta_{y0})$ 
of Eq.~(\ref{eq:GaussMode}) with $\theta_y = 0$ (upper insets) 
and $\theta_x = 0$ (right insets), while $\theta_{y0}=0$ and  
$\theta_{x0}$ as per cases (a) and (b).}
\label{fig:ak}
\end{figure}
%%%%%%%%%%%%%%%%%%%%%%%%%%%%%%%%%%%%%%%

In Fig. \ref{fig:ak} we show the angular probability distribution 
of the emitted photon. 
The beam divergence $\Delta \theta_x = \frac{1}{k_0 \Delta x} \simeq 0.015 \pi$ 
in the $x-z$ plane is almost twice larger than that 
$\Delta \theta_y = \frac{1}{k_0 \sigma_y} \simeq 0.008 \pi$ in the $y-z$ plane, 
consistent with the narrower spatial distribution 
$\Delta x \simeq 2.6\:\mu\textrm{m} < \sigma_x = 4\:\mu\textrm{m} $ 
of the atomic excitation (or emission) probability $P(\bos{r})$, 
as discussed above. 
In the collinear geometry, $\bos{k}_p, \bos{k}_d \parallel \hat{z}$,
the radiation is emitted at a small angle 
$\theta_{x0} = \bos{k} \protect\angle \hat{z} \simeq 0.014 \pi$ 
due to the detuning induced phase gradient 
of the atomic amplitudes $b_j$ along $x$. 
With a small angle $\bos{k}_d \protect\angle \bos{k}_p = 0.009 \pi$ 
between the drive and the pump fields, the latter still propagating 
along $z$, we can compensate this phase gradient, 
resulting in the photon emission along $z$ ($\theta_{x0}=0$).
We may approximate the angular profile of the emitted radiation 
with a Gaussian function
\begin{equation}
B_{\bos{k}} \propto B(\theta_x,\theta_{x0};\theta_y,\theta_{y0}) = 
e^{-(\theta_x - \theta_{x0})^2/\De \theta_x^2}
e^{-(\theta_y - \theta_{y0})^2/\De \theta_y^2}. \label{eq:GaussMode}
\end{equation}
We then see from Fig. \ref{fig:ak}  that in the $y-z$ plane the angular 
profile corresponds to a Gaussian mode with $\theta_{y0} = 0$, 
but in the $x-z$ plane the angular profile deviates from the Gaussian,
the more so for the case of the corrected emission angle $\theta_{x0} = 0$.
To fully collect this radiation, we thus need to engineer an elliptic
lens with appropriate non-circular curvature along the $x$ direction.

The total probability of radiation emitted into the free-space spatial 
mode $\mc{E}(\bos{r})$ subtending the solid angle 
$\Delta \Omega = \pi \Delta \theta_x \Delta \theta_y$ is 
$P_{\Delta \Omega} \simeq 0.74$.
%\texttt{[David: This is coincidentally close to that in our 
%PRL 121, 123605 (2018), where we have $N=1000$ with $\sigma_{x,y} = 1\:\mu$m. 
%But here we took $\sigma_{x,y} = 4\:\mu$m, i.e. 16 times the cross section 
%and thus smaller solid angle, and $N=15000$, i.e. 15 times the atom number,
%so we get about the same ;).]} 
This probability can be increased by optimizing the geometry of the sample, 
e.g., making it narrower and longer, as discussed above. 
Alternatively, we can enhance the collection efficiency of the coherently 
emitted radiation by surrounding the atoms by a moderate finesse, one-sided 
optical cavity. Assuming a resonant cavity with frequency $\om_{k_0}$, 
mode function $u_o(\bos{r})$ and length $L_o$, the overlap 
$v = \frac{1}{\sqrt{L_o}} \int d^3 r \mc{E}(\bos{r}) u_o^*(\bos{r})$
determines the fraction of the radiation emitted by the atomic
ensemble into the cavity mode, while the cavity finesse $F$ determines 
the number of round trips of the radiation, $n \simeq F/2\pi$, and thereby 
the number of times it interacts with the atoms, before it escapes the cavity. 
The probability of coherent emission of radiation by $N$ atoms into the 
cavity output mode is then 
$P_{\mathrm{out}} \simeq \frac{|v|^2 n N}{|v|^2 n N +4\pi}$. 

\section{Conclusions}
 
We have proposed a scheme for coherent microwave to optical conversion 
of a photon of a superconducting resonator using an ensemble of atoms 
trapped on a superconducting atom chip. 
The converted optical photon with tailored temporal and spatial profiles
can be fed into a waveguide and sent to a distant location, where the 
reverse process in a compatible physical setup can coherently convert 
it back into a microwave photon and, e.g., map it onto a superconducting qubit. 

In our scheme, the atoms collectively interact with the microwave cavity 
via a strong, dipole-allowed Rydberg transition. We have considered the
conversion of at most one microwave photon to an optical photon, 
for which the interatomic Rydberg-Rydberg interactions are absent. 
In the case of multiple photons, however, the long-range interatomic 
interactions will induce strong non-linearities accompanied by
the suppression of multiple Rydberg excitations within the blockade 
volume associated with each photon \cite{Murray2016,Firstenberg2016}. 
This can potentially hinder the microwave photon conversion 
and optical photon collection due to distortion of the temporal 
and spatial profile of the emitted radiation. 
%Our scheme, however,
%is primarily intended for communication between microwave operated 
%quantum sub-registers and, as such, it should ideally deal with 
%at most one microwave photon encoding a qubit state at a time. 
%Yet, in the limit of a few photons interacting with many atoms 
%in a volume much larger than the blockade volume, we expect 
%the conversion to be well approximated as a linear process. 

\begin{acknowledgments}
We acknowledge support by the US ARL-CDQI program through cooperative agreement
W911NF-15-2-0061, and by the DFG SPP 1929 GiRyd and DFG Project No. 394243350. 
D.P. is partially supported by the “HELLAS-CH” (MIS Grant No. 5002735),
%implemented under ``Action for Strengthening Research and Innovation 
%Infrastructures,'' funded by the Operational Programme ``Competitiveness, 
%Entrepreneurship and Innovation'' (NSRF 2014-2020) and co-financed by Greece 
%and the European Union (European Regional Development Fund)), 
and is grateful to the Aarhus Institute of Advanced Studies for hospitality, 
and to the Alexander von Humboldt Foundation for additional support 
in the framework of the Research Group Linkage Programme.   
\end{acknowledgments}

\end{document}